# Hosting Byzantine Fault Tolerant Services on a Chord Ring

Alan Dearle, Graham NC Kirby, Stuart J Norcross
*School of Computer Science*
*University of St Andrews*
*St Andrews*
*Fife*
*Scotland*
*al@cs.st-and.ac.uk, graham@cs.st-and.ac.uk, stuart@cs.st-and.ac.uk*

## Abstract

*In this paper we demonstrate how stateful Byzantine Fault Tolerant services may be hosted on a Chord ring. The strategy presented is fourfold: firstly a replication scheme that dissociates the maintenance of replicated service state from ring recovery is developed. Secondly, clients of the ring based services are made replication aware. Thirdly, a consensus protocol is introduced that supports the serialization of updates. Finally Byzantine fault tolerant replication protocols are developed that ensure the integrity of service data hosted on the ring.*

## 1. Introduction

In this paper we demonstrate how stateful Byzantine Fault Tolerant (BFT) [1] services may be provided on a Chord ring.

The approach is to implement a service as a number of service components, each of which is hosted on a node of the Chord network. A mapping is established between the entities over which the service operates and the key space of the peer-to-peer overlay. A service component is responsible for the entities whose keys route to the overlay host. Thus the ability to map from service parameters to a key and hence to a ring node hosting a service component responsible for that key is a requirement of this technique.

Service components are discovered using standard key based routing protocols and, once discovered, clients interact directly with the service components to obtain the desired service.

This paper makes several contributions. Firstly, we demonstrate the problems that arise when the strategies used to maintain Chord ring integrity are applied to the service components hosted on the ring infrastructure. We present an alternative strategy which may be used to provide Byzantine Fault Tolerant services on a Chord ring. We illustrate this by demonstrating how a BFT distributed hash table (DHT) with update may be provided on a standard non-BFT Chord infrastructure.

Our approach is fourfold: firstly a replication scheme that dissociates the maintenance of replicated service state from ring recovery is developed. Secondly, clients of the ring based services are made replication aware. Thirdly, a consensus protocol is introduced that supports the serialization of updates. Finally Byzantine fault tolerant replication protocols are developed that ensure the integrity of the service data hosted on the ring.

## 2. Background

A number of P2P overlay protocols have been proposed that support the Key-based Routing (KBR) abstraction [2-5]. Under a KBR scheme each addressable application level entity has an associated key value and each key value maps to a unique live node in the overlay network. Upcalls from the routing layer inform the application layers of changes to the keyspace, thus allowing an application to become aware of changes to the set of keys that map to the local node.

Each scheme provides an overlay structure that links a participating node to a small number of peer nodes with which it can communicate. Each of these systems provides routing mechanisms enabling nodes to be addressed using a *key* value in *log n* time, where *n* is the number of nodes. The P2P architectures are all self-repairing in the face of host or network failures[1]. The different overlay mechanisms differ considerably in the way in which the routing algorithms are implemented. However, the different systems may be usefully be classified as being in one of three families: Chord-like systems [2]; Plaxton-like systems [3]; and CAN-like systems [4]. These systems offer a variety of abstractions [10] built on or related to the KBR abstraction. Those most closely related to this work are the *Distributed Object Location and Routing* (DOLR) [6], and *group anycast/multicast* (CAST) abstractions.

[1] Up to some limit governed by the frequency of failures and the amount of state maintained by the nodes.

The DOLR abstraction is concerned with the implementation of a decentralised discovery service in which applications may place objects on arbitrary nodes within an overlay and announce their existence using a key. It exports two operations: *publish* and *sendToObject*. The former is used to publish the association of an object with some key. The *sendToObject* operation causes a message to be sent to a number of copies of the object(s) with a specified key. The CAST abstraction is used to implement multicast groups. In its simplest form, it exports two operations *join* and *cast*. The *join* operation permits a node to join a multicast group specified by a key as a parameter. This causes a message to be routed to the node responsible for that key. Whenever a node is encountered that is already a member of the group, the node is added as a child of that node. Thus a multicast tree is formed, rooted at the node responsible for the specified group key. When a *cast* call is made, a message is sent to the root for the key. The root instructs its children to send the message to the nodes in their dissemination tree. This process repeats recursively until all the nodes in the group have been sent the message.

## 2.1 Chord

In this paper we focus on one particular peer-to-peer routing protocol – Chord. Chord is a ring based protocol that supports KBR. At its most basic level, Chord only requires each node to maintain a pointer to its immediate successor in the ring. Each node $N$ is assigned a unique *m-bit* identifier key $K_N$ and the ring is arranged in key order where keys are ordered on an identifier circle modulo $2^m$. Every key value maps to a unique live node in the overlay network.

The Chord protocol supports a *lookup* operation which takes a key value and returns the network address of the Chord node in the overlay network to which the key value maps. Each node $N$ is responsible for the region $[K_{pred(N)}, K_N)$ where $K_N$ is the node's key and $K_{pred(N)}$ that of its predecessor. Thus, a lookup on key $k$ will yield the address of the node $N$ whose key $K_N$ is the first key in the ring that is equal to or greater than $k$ in the keyspace (modulo keyspace). In this way, the Chord protocol provides a lookup service mapping keys to overlay nodes. We call the Chord node $N$ that is returned by Chord's *lookup* method when called with key $k$ the *primary node* for $k$.

In order to guarantee correct lookups each node need only know its correct successor and as such the lookup request can be passed around the ring until the appropriate node is found. With such a simple scheme lookup times vary linearly with the number of nodes in the ring. To improve lookup times each node maintains additional routing state called the *finger table* which contains up to $m$ entries. From [2]:

> *"The $i^{th}$ entry in the table at node N contains the identity of the first node S that succeeds N by at least $2^{i-1}$ on the identifier circle,…"*

The *finger table* is consulted during the iterative lookup process in which, at each stage, the node referenced from the current node and with the closest preceding key to the desired key is chosen to be used in the next stage of the iteration. This reduces lookup time to O(log $X$) where $X$ is the number of nodes in the ring. To support self-repair of the ring, each node also maintains a successor list of $k$ nodes which immediately follow the node in the ring order.

The successor list permits a node to find its new correct successor should its successor fail. For a successor list of size $k$ the system is resilient to up to $k$-$1$ successive nodes failing within a given interval. This provides resiliency of the ring and the look-up protocol, but does not ensure the integrity of the data structures hosted by ring nodes. We will return to this issue momentarily.

Figure 1 shows a simple Chord ring. Each node contains references to its successor (single filled arrow head) and its predecessor (dashed line and open arrowhead). Node 1 has been elaborated to show its successor list (double headed arrows) and its fingers (bold chords across ring).

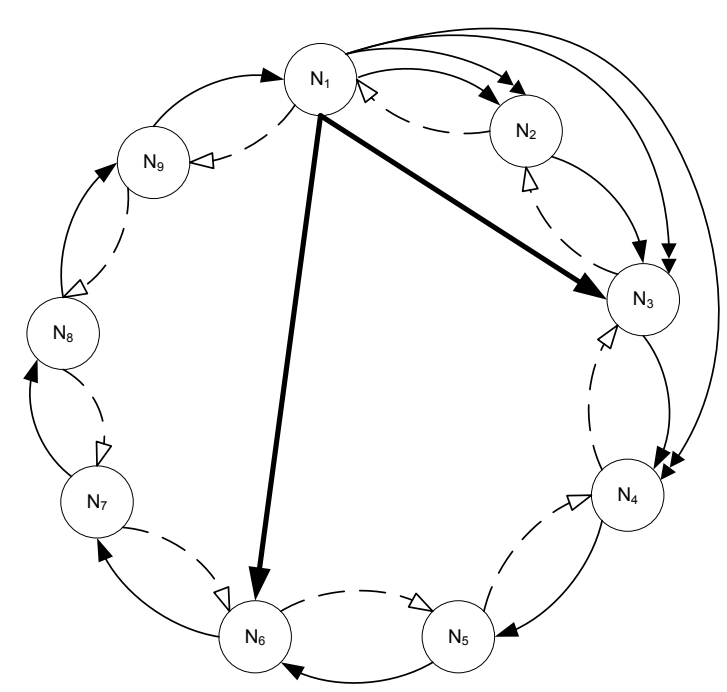

**Figure 1**

## 2.2 Providing Services on Chord

In this paper we demonstrate how a distributed service may be implemented on a Chord ring. To illustrate our approach, we focus on an updatable distributed hash table (DHT) service mapping keys to values. This simple service has sufficient attributes to illustrate the more general approach. In particular, it requires serialization of updates.

The approach is to co-locate the service component responsible for some key range on the primary node responsible for that range. In order to make the global service resilient to failure, it is necessary to replicate the state of the service components that implement it. The obvious approach is to co-locate the replicas on the successors of the primary. However, this simple approach of conflating the resilience of service components hosted on the ring with the resilience of the ring itself can be dangerous.

To understand why this is so, some examination of the Chord protocol is necessary. Using Chord, each node $N$ is responsible for the region of keyspace $[K_{pred(N)}, K_N)$, that is the keyspace up to its own key and following the key of its predecessor. Consider a lookup starting at Node $N_1$ of key $k$ where $K_5 < k < K_6$. In this case, as described in [2], the lookup algorithm visits the following nodes: $N_3$, $N_4$, $N_5$ and returns $N_6$. Note that although node $N_1$ has a finger table entry referring to the hosting node, $N_6$, it cannot be used since the $K_6$ is not less than the key being searched for and to use it would (always) risk overshooting the target. Due to this and in general, all Chord lookup operations are always routed via the predecessor of the node responsible a given key.

In the event of a failure of node $N_1$ in Figure 1, the ring will self repair using the Chord repair protocols and node $N_2$ will become the successor of node $N_9$. However, consider the case in which node $N_1$ starts to operate incorrectly either maliciously or simply erroneously. Since no ring failure has occurred, the successor and predecessor references will remain as shown in Figure 1 and all lookup operations for keys in the range $[K_8, K_1)$ will be routed to $N_1$. Furthermore, any attempts to access the successor list of $N_1$ will also be routed via $N_1$ since other nodes elsewhere in the ring have no knowledge of the node topology or key space in the vicinity of node $N_1$. Of course, the successors of node $N_1$ are mostly in the successor list of $N_9$, however, they are not used for addressing unless $N_1$ is known to be faulty which, in general, it cannot be assumed to be. Furthermore, a node may operate correctly at the P2P level and erroneously at the service level. Thus, using standard Chord protocols, a single erroneous node in the ring can prevent access to the services for which it is responsible both on the primary and on its replicas.

## 3. Dissociating Replicated Service State Maintenance from Ring Recovery

Consider the Chord ring shown in Figure 2. For brevity no finger tables, predecessors, successors or successor lists have been shown. Node $N_1$ at the top of the figure has key $K_1$ and is responsible for the range $[K_{15}, K_1)$ which is shown by the dark gray segment. Our strategy for dissociating the maintenance of replicated service state from the ring infrastructure is to replicate that state on nodes located around the ring. For a replication factor of $r$, the state associated with key $k$ is replicated on $r$–1 nodes associated with keys $k + nKS/r$ where $n$ ranges from 1 to $r$–1 and $KS$ is the size of the keyspace. For a replication factor of four, the regions of keyspace corresponding to the replicas of $N_1$'s keyspace are shown in light gray and labeled $R_1$, $R_2$ and $R_3$. For a given key $k$ we call the nodes responsible for keys $\{k, k + KS/r, k + 2KS/r\ldots\}$ the *peer set* of *k.*

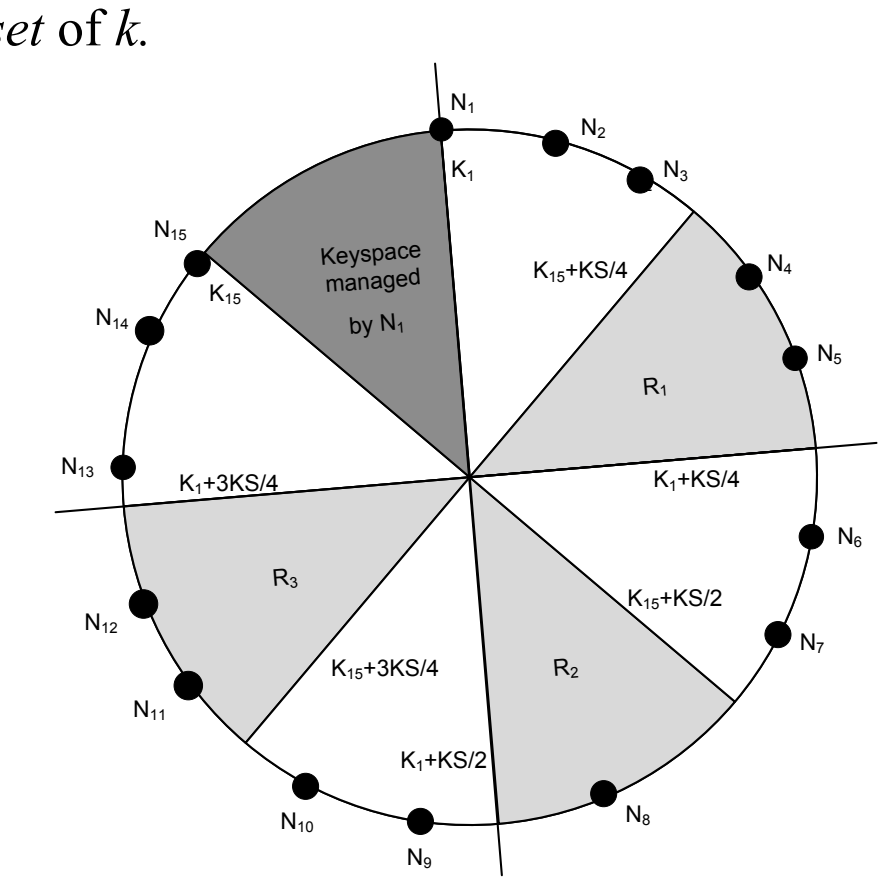

**Figure 2**

In Figure 2, it is clear that replica key ranges do not, in general, map onto single nodes. For example, replica range $R_1$ is part-owned by three nodes, $N_4$, $N_5$ and $N_6$. Similarly $R_3$ is part-owned by $N_{11}$, $N_{12}$ and $N_{13}$. Thus the replica state of a node is not found in its entirety on ($r$-1) other nodes; instead, it is spread through a collection of nodes. However the state corresponding to a given key $k$ may be always found on exactly ($r$-1) replica nodes barring failures. Thus, as shown in Figure 2, replicas of (part of) the service state hosted by node $N_1$ are stored on nodes $N_4$, $N_5$, $N_6$, $N_8$, $N_9$, $N_{11}$, $N_{12}$ and $N_{13}$. We call the key ranges $R_1$, $R_2$ and $R_3$ *peer key ranges*. This does not add complexity to client discovery of replicas but does impact on the complexity of the fault tolerant protocols that maintain the replicated state. The service component hosted by a node typically only supports interaction with entities whose keys map to that node, e.g. the service component on node $N_1$ is associated with entities with keys in the range $[K_{15}, K_1)$. We refer to such a key as the *natural key* of an entity. Without considering replication, keys in this range would not normally be stored on the replica nodes. Two different approaches may be taken to replicating service component state on replica nodes:

- Associate service component replica state with natural keys, and,
- Associate service component replica state with calculated replica keys.

The first of these approaches means that service components are associated with replica nodes with keys outwith the key range managed by the replica. For example, in Figure 2, replica state corresponding to key $K_1$ might be associated with node $N_6$. By contrast, using the second scheme, each replica state is associated with a key calculated by shifting the original key by an appropriate fraction of the keyspace. In Figure 2 the replica of the state associated with key $K_1$ would be associated with key $K_1 + KS/4$ on Node $N_6$ which is in the normal key range managed by that node. In general, the key is calculated by taking the natural key of the service and adding $nKS/r$ to it. Using the first approach server-side checks need to be relaxed to permit state to be stored that corresponds to the peer key ranges. By contrast, if the second option is chosen, no such relaxation is necessary. In our implementations we always associate replica state with a calculated key since it makes server code less complex.

## 4. Making Clients Replication Aware

Some BFT approaches, notably that of Castro and Liskov [7], require a primary to be identified which coordinates the protocols. Having a primary makes the serialization of operations simpler but adds complexity in that election protocols for primaries are required in the face of failure. We believe that a symmetric scheme is simpler overall by avoiding the need for election protocols and does not rely on the immediate detection of a Byzantine primary. Our algorithms remove the need for a primary by making the client replica aware. Using the scheme described in this paper, each client needs to be aware of the replication factor and needs to be able to independently address ring nodes.

We sketch the algorithms used for the two primary DHT operations – *put* and *get*. As is traditional a *put(key, value)* will update the value associated with some key and a *get(key)* will return that value. As described in [1], we require at least $3f+1$ replicas to provide BFT in which up to $f$ replicas are faulty (since it must be possible to complete operations after communicating with $n–f$ replicas). In this paper, to simplify explanation, we assume the simplest case of at most 1 faulty node and thus a system with four copies of service component state. In our implementations we follow a generative approach permitting a (statically determined) arbitrary number of faulty nodes to be tolerated.

The general approach followed is that clients send requests to all peers in the peer set. Each of the replica peers is discovered by routing to the node responsible for the corresponding replica key; such routing may be made via an arbitrary ring node. This obviates the possibility that one faulty node may prevent the discovery of replica nodes and hence the operation from being carried out. The client waits for replies from the replica nodes and when an appropriate number of consistent replies are received the operation is considered to be complete. It is sufficient for the client to receive $f+1$ consistent replies since at most $f$ nodes may be faulty.

The algorithms for *put* and *get* are similar in nature. For brevity we only show the pseudo code for *put* in Figure 3. The client calculates the set of keys to which data must be written and attempts to store the data on the appropriate nodes. Routing to nodes may fail as may individual nodes and so the process is repeated until an appropriate number of *put*s have been made on the replica nodes according to BFT assumptions. From the client perspective, it appears that no server coordination is being performed. However, as discussed in the next section, this is not always the case.

```
UID ClientPut(Data data) throws Exception {
   int MAX_FAULTY = floor((REP_FACTOR - 1)/3)
   UID uid= hash(data)
   Set<Key> keys_not_stored =
         calculateKeys(uid)
   while size(keys_not_stored)>MAX_FAULTY {
      parforeach Key k in keys_not_stored {
      try {
         Server node = routeTo(k)
         node.put(data)
         remove k from keys_not_stored
      }
      catch(TimeoutException,CommsException) {
         // try again – up to policy limit
      }
   }
   return uid
}
```

**Figure 3**

## 5. BFT Consensus Protocol

Since the algorithms do not have a primary node, a consensus protocol is needed in cases where a serial ordering of operations is important. This is the case with the *put* operation which is an update operation and subject to race conditions. To satisfy this requirement, we have developed a consensus protocol which is essentially a counting algorithm. Space prohibits the algorithm from being described in full and we will sketch out the mechanisms here; further details may be found at [8].

In the algorithm sketched below, each server interacts with the other servers in its peer set that are responsible for a specified unique id *uid*. The peer set members are calculated using the address arithmetic described in Section 2. The underlying routing protocols provided by Chord are used for discovering these nodes.

At a high level, a two-phase algorithm is executed on each server. The phases involve counting both votes and commits for an update request. Each phase completes when the BFT message thresholds[2] have been received. The first phase is initiated by the receipt of a *put* message on a server. However, this phase is only entered into if the server is not already engaged in an update of the specified *uid*. If it is, the request is queued until the previous update has been completed successfully or otherwise. Each server maintains a per *uid* state machine which records the following information: if a put has been received, a count of votes received, whether a vote has been sent, a count of commits received, whether a commit has been sent, whether the node is already engaged in a put, and, whether the server has chosen the update to which the state machine pertains.

The state machine is relatively complex with 33 states necessary to capture the asynchrony in the four way replication scheme shown in Figure 2. However the algorithm is conceptually simple: each node communicates with its peer set sending *vote* messages in response to the receipt of a *put* from a client. When the BFT threshold of votes has been received by a server, a *commit* message is sent to its peer set. When enough *commit* messages have been received from other peers, the transaction is made and the client is notified.

Consistent serialization is achieved by allowing an update voted for by a sufficiently high number of other servers to proceed ahead of a previous locally selected update. Since there is no guarantee that any one of a set of concurrent updates will gain enough votes to reach this threshold, the algorithm may deadlock. This may be handled by a timeout/retry mechanism with a randomized backoff.

[2] The BFT thresholds are in fact different for vote and commit messages – we require $2f+1$ vote messages and $f+1$ commit messages.

## 6. BFT Node Recovery

The final mechanism needed to provide BFT is a recovery mechanism to ensure the integrity of service component state when the underlying ring changes. There are two categories of change that must be accommodated: (a) nodes joining and (b) nodes leaving the ring; the latter may be orderly or due to failure. When a new node is added to a ring, the effect is to reduce the keyspace of the new node's successor. Conversely, a node leaving a ring causes the keyspace of the leaving node's successor to be increased. When the topology changes, the standard Chord protocol provides upcalls from the P2P routing layer to notify the software hosted on nodes of a change of ring topology. In both cases (a) and (b) above, the upcall mechanism provides the service layer with the old and new key ranges for which the node is responsible and initiates the algorithms that repair the services hosted on the ring. In practice it is useful to separate the upcalls, and in our implementations two different upcalls are used to start the repair process: *release(old_range)* and *takeOver(extra_range)*.

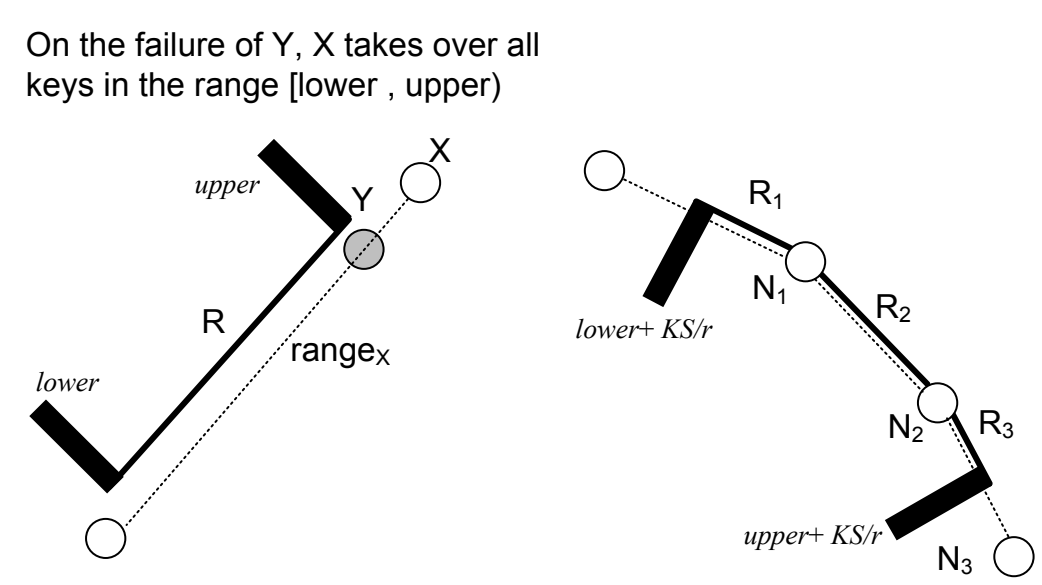

**Figure 4**

Again, space precludes a full exploration of the algorithms that are invoked in response to these upcalls; they will therefore be sketched out here. The difficult case is when a node fails and we therefore concentrate on that case. When a *takeOver* upcall is received, the algorithm calculates the address ranges of the (non-failing) nodes in the peer set. Consider the example shown in Figure 4 in which node Y has failed and node X has taken over the key range R=[*lower*, *upper*). In order to recover, X needs to obtain all service component data in range R from the peerset of the range. This task is slightly more complex that it might seem since (a) X has no knowledge of what extant keys lie in this range; and (b), the replica ranges of R will, with high probability, be split between multiple replica nodes for each of the regions { [*lower*+ *KS/r*, *upper*+ *KS/r*), [*lower*+ 2*KS/r*, *upper*+ 2*KS/r*)… } as shown in Figure 4 with nodes $N_1$, $N_2$ and $N_3$ which all manage replica keys in the range [*lower*+

*KS/r*, *upper*+ *KS/r*). Thus the recovery process is multi-phased. First a set of the *peer servers* holding replica data is constructed by repeatedly routing to the lowest key in each of the replica key ranges and following their successor links. Next, each of the peer servers is requested to return the set of keys they hold. By counting and matching replies, a list of keys may be constructed on the node performing the reconstruction. This set will contain keys in the range R and from the corresponding peer sets of R. Finally, the data corresponding to these keys is asynchronously fetched from enough replica servers to be safe under BFT assumptions.

Within this algorithm there are many subtleties which have been glossed over here for brevity. These include: the policy choice of which servers from which to fetch data, the exploitation of self-verifying data to avoid multiple fetches, the avoidance of fetching more data than is required and dealing with transient failures. A final complexity is that if the ring is not stable, not all the nodes that may be requested for data will actually have it.

## 7. Summary and Conclusions

The techniques demonstrated in this paper make several novel contributions. Primarily we demonstrate that a Chord-like P2P system may be used to host Byzantine Fault Tolerant services. The techniques described are applicable in other KBR systems although the specific problems of primary node failure are not as critical in other systems, for example, those based on Plaxton routing. In many ways, the dissociation of the maintenance of replicated service state from the underlying KBR mechanism increases the applicability of the techniques. Making clients aware of the addressing mechanisms used to address nodes is a critical element in the establishment of BFT mechanisms. The consensus protocol used to support the serialization of updates is an optional part of the scheme that may be applied when serialization is required.

The algorithms sketched in this paper have all been implemented as part of the Autonomic Storage Architecture (ASA) project. The state machine corresponding to the consensus protocol described in Section 5 is dependent on the replication factor used. We have therefore applied generative techniques to automatically generate state machines for a given replication factor.

## 8. Ongoing work

The algorithms sketched in this paper are being applied in the ASA project which is constructing a distributed autonomic file system. In particular, the BFT protocols are being applied to the management of data structures that maintain mappings from globally unique file identifiers to sequences of file versions. We also currently investigating programming language constructs which present resilient service abstractions implemented by the mechanisms described in this paper.